%% file: main.tex
\DocumentMetadata{}
\documentclass[10pt, sigconf]{acmart}

\usepackage[utf8]{inputenc}
\usepackage{tikz}
\usepackage{listings}
\usepackage{amsmath}
\usepackage{lipsum}
\usepackage{mathtools}
\usepackage{cuted}
\usepackage{pifont}
\usepackage{amsfonts}
\usepackage{relsize}
\usepackage{booktabs}
\usepackage[super]{nth}
\usepackage[noend]{algpseudocode}
\usepackage[ruled, vlined, linesnumbered, noend]{algorithm2e}
\usepackage{afterpage}
\usepackage{tablefootnote}
\usepackage{xspace}
\usepackage[most]{tcolorbox}
\setcopyright{none}
\renewcommand\footnotetextcopyrightpermission[1]{} % removes footnote with conference information in first column
\renewcommand{\i}{\textit{(i)}~}
\newcommand{\ii}{\textit{(ii)}~}

\usepackage{subcaption}

\newcommand{\namex}{Themis\xspace}

\title{A Tale of Two Scales: Reconciling Horizontal and Vertical Scaling for Inference Serving Systems}
\author{Kamran Razavi}
\authornote{Equal Contribution}
\affiliation{%
 \institution{Paderborn University\\ Technical University of Darmstadt}
}
\author{Mehran Salmani}
\authornotemark[1]
\affiliation{%
 \institution{Ilmenau University of Technology}
}
\author{Max M\"uhlh\"auser}
\affiliation{%
 \institution{Technical University of Darmstadt}
}
\author{Boris Koldehofe}
\affiliation{%
 \institution{Ilmenau University of Technology}
}
\author{Lin Wang}
\affiliation{%
 \institution{Paderborn University}
}
\begin{document}

% \keywords{Inference Serving Systems, Autoscaling, Machine Learning}

\input{sections/abstract}
\maketitle
\thispagestyle{plain}
\pagestyle{plain}
\input{sections/sec_1_introduction.tex}
\input{sections/sec_2_motivation.tex}
\input{sections/sec_3_system_design.tex}
\input{sections/sec_4_problem_statement.tex}
\input{sections/sec_5_transition.tex}
\input{sections/sec_6_experimental_results.tex}
\input{sections/sec_7_related_works.tex}

\input{sections/sec_8_conclusion.tex}
\input{sections/acknowledgement}

\clearpage

\bibliographystyle{ACM-Reference-Format}
\bibliography{ref}
% \clearpage
% \input{sections/appendix}

\end{document}

%% file: sections/abstract.tex
\begin{abstract}
Inference serving is of great importance in deploying machine learning models in real-world applications, ensuring efficient processing and quick responses to inference requests. However, managing resources in these systems poses significant challenges, particularly in maintaining performance under varying and unpredictable workloads. Two primary scaling strategies, horizontal and vertical scaling, offer different advantages and limitations. Horizontal scaling adds more instances to handle increased loads but can suffer from cold start issues and increased management complexity. Vertical scaling boosts the capacity of existing instances, allowing for quicker responses but is limited by hardware and model parallelization capabilities.

This paper introduces \namex{}, a system designed to leverage the benefits of both horizontal and vertical scaling in inference serving systems. \namex{} employs a two-stage autoscaling strategy: initially using in-place vertical scaling to handle workload surges and then switching to horizontal scaling to optimize resource efficiency once the workload stabilizes. The system profiles the processing latency of deep learning models, calculates queuing delays, and employs different dynamic programming algorithms to solve the joint horizontal and vertical scaling problem optimally based on the workload situation. 
Extensive evaluations with real-world workload traces demonstrate over $10\times$ SLO violation reduction compared to the state-of-the-art horizontal or vertical autoscaling approaches while maintaining resource efficiency when the workload is stable. 

% \namex{} aims to provide a responsive, resource-efficient solution that ensures end-to-end inference latency while reducing operational costs. 
% The paper details the design and architecture of \namex{}, presents a problem formulation based on integer program, solves it by two different dynamic programming algorithms based on the workload situation, and evaluates the system with real-world workload traces, demonstrating over $10\times$ SLO violation reduction compared to the state-of-the-art horizontal or vertical autoscaling approaches. 
\end{abstract}

%% file: sections/sec_1_introduction.tex
\section{Introduction} \label{sec:intro}
\iffalse
\begin{itemize}
    \item what is the domain (DL) with an example why this is important
    \item what are the challenges (resource efficiency , SLA, dynamic workload, dependency) with a scenario
    \item why the challenges cannot be solved easily (a bit of related work)
    \item opportunities leading to the solution
    \item the solution + the approach
    \item contribution + structure of the paper
\end{itemize}
\fi
Cloud-based deep learning (DL) inference is a common component in modern intelligent applications and services, often involving multiple DL models interconnected in a dataflow multi-model (pipeline) system~\cite{razavi2022fa2}. An example is a real-time video analytics application for traffic management, which may include models for video frame extraction, object detection, object classification, and tracking~\cite{elhoseny2020multi}. The performance of such systems is mostly evaluated based on two indicators: user satisfaction, quantified through Service Level Objectives (SLOs) mainly on the end-to-end latency~\cite{gujarati2017swayam}, and resource efficiency~\cite{zhang2019mark}, referring to the optimal utilization of computational resources. These performance indicators ensure the delivery of high-quality results while efficiently using resources, which is crucial for the scalability and sustainability of cloud-based DL applications~\cite{luo2022power, razavi2022fa2, zhang2019mark, gunasekaran2022cocktail, ali2020batch, mounesan2024reinforcement}.

Efficient resource management in DL inference serving systems is crucial to maintaining system performance, especially under variable and unpredictable workloads~\cite{razavi2024sponge, salmani2023reconciling}. 
Two primary scaling strategies are often used to manage these dynamic workloads: horizontal and vertical scaling. 
Horizontal scaling involves the addition of more instances of the same DL model to handle increased loads, facilitating workload distribution without altering the resource configuration of existing instances. 
By distributing the workload across several instances, horizontal scaling can handle large workloads and maintain performance as demands grow.
However, bringing up new instances in horizontal scaling involves cold start issues~\cite{lemay2020perseus, luo2022power, pavlenko2024vertically, romero2019infaas} due to booting up additional instances, configuring them, and joining them into the system, which can take several seconds to minutes. 
The cold start issue reduces system responsiveness when the workload changes unpredictably, 
% resulting in 
leading to SLO violations and 
% decreasing 
reduced user satisfaction.
Additionally, horizontal scaling increases system management's complexity, requiring more sophisticated load balancing.

Vertical scaling, conversely, focuses on adding resources such as CPU cores to existing DL model instances, thereby boosting the capacity of existing instances to handle more tasks simultaneously. 
A recently announced feature named in-place vertical scaling~\cite{kubernetes_in_place} by Kubernetes~\cite{kubernetes}---a dominant open source system to orchestrate and manage the containers in the system---allows additional resources
% (such as CPU) 
to be added to existing instances without the need for rebooting or downtime, 
% providing 
offering a significant advantage 
% compared to 
concerning the cold start issue. Additional computing resources, such as 
% multiple 
extra CPU cores, allow inference requests to 
be batched and processed 
%the parallelizable tasks to be processed 
concurrently, 
% reducing batch execution time. 
improving the system throughput. 
This capability enables a much quicker response to increased load compared to setting up new instances as required in horizontal scaling.
While vertical scaling can provide rapid speed-up by instantly increasing the resources of a single instance, it is often limited by the physical capabilities of the hardware and the parallelizability of the DL model.
% limitation. 
Once those limits are reached, no further speed-up is possible, thus reaching the maximum throughput. In contrast, horizontal scaling can continue to expand by adding more instances.
Furthermore, vertical scaling can lead to higher costs due to the need for high-end hardware, 
% upgrades, 
and it presents a single point of failure risk, where if the 
% upgraded 
single instance fails, it can significantly impact system stability.
% performance.
% Both strategies offer critical flexibility and scalability in managing dynamic workloads; they can be used to satisfy the stringent SLOs related to latency and system performance.

Existing inference serving systems mainly focus on horizontal scaling mainly due to the benefits of the scalability of horizontal scaling and the limitations of vertical scaling.  
% the limitations of vertical scaling and the benefit of having a higher throughput in horizontal scaling since DL models are not 100\% parallelizable, thereby cannot fully utilize extra resources provided by vertical scaling. 
Studies like INFaaS~\cite{romero2019infaas} focus on individual DL models where horizontal scaling is used in response to workload changes by bringing up new virtual machines. Others have considered DL pipelines and use horizontal scaling to 
% save on 
optimize resource costs by switching between hardware types or model variants when the workload changes~\cite{2020-socc-inferline, ghafouri2024ipa, razavi2022fa2}.
% INFaaS~\cite{romero2019infaas} uses horizontal scaling for individual DL models model in response to workload changes by bringing up new virtual machines. InferLine~\cite{2020-socc-inferline} uses horizontal scaling for DL pipelines to save on costs by switching between hardware types. IPA~\cite{ghafouri2024ipa} reduces operational costs while maintaining user satisfaction by horizontal scaling and incorporating DL model variants. FA2~\cite{razavi2022fa2} guarantees requests end-to-end latency by designing a new horizontal autoscaler and solving data dependency, uncertainty, and dynamic batching problems. 
All of the mentioned approaches suffer from cold start issues when facing dramatic workload increases. Sponge~\cite{razavi2024sponge} uses in-place vertical scaling and changes the CPU allocation of the running instance to absorb the sudden changes in the workload. However, they only consider a single DL model and 
% miss 
ignore the challenges behind the data dependency of pipeline inference serving systems. Moreover, their approach suffers from SLO violations 
% in the case of a high request rate workload 
under high request rates that are 
beyond the capabilities of one powerful instance due to parallelization and physical hardware limitations, as discussed above. 
In short, 
% current autoscalers 
existing autoscaling solutions suffer from non-responsiveness under 
% sudden workload changes
high workload variability, and to compensate for that, they require a significant 
% degree 
level of over-provisioning. 
% Therefore, an ideal autoscaler 
Ideal autoscaling should be responsive, meaning it can immediately react to sudden workload changes, and 
% . It should also be 
resource-efficient, avoiding over-provisioning, 
% and prevent 
while limiting SLO violations by ensuring end-to-end inference latency. 
% To meet these requirements, the ideal autoscaler should switch between horizontal and vertical scaling strategies, leveraging the unique benefits of each approach.
% and they do not consider horizontal scaling.

In this work, we explore how to reduce SLO violations and resource costs by studying an autoscaling mechanism that combines both vertical and horizontal scaling mechanisms and leverages the benefits of both 
% horizontal and vertical scaling 
mechanisms in the pipeline inference serving system. Based on the insights from such an exploration, we design an autoscaler, named \namex{}, that uses a two-stage autoscaling strategy, where it first leverages the responsiveness characteristics of in-place vertical scaling 
% characteristics 
to absorb the 
% unexpected 
unpredictable surge of requests in the workload. Second, \namex{} performs horizontal scaling to optimize resource 
% consumption 
efficiency when the workload 
% becomes stable 
tends to stabilize by 
% bringing more 
transitioning to a set of less powerful instances. 
% We later show that horizontal scaling provides a higher throughput compared to vertical scaling under the same amount of resources.

To be precise for the vertical scaling strategy, after getting the application pipeline with its SLO, \namex{} 
% solves 
addresses the challenges regarding the in-place vertical scaling decision-making in pipelines by first profiling the processing latency of all the DL models with respect to different CPU and batch size allocations, and calculating the queuing delay based on the given batch size and the arrival rate. Next, \namex{} finds the right amount of resources to serve the incoming workload (avoiding under- or over-provisioning) for all the models in the system, in a single shot, by encapsulating the autoscaling problem in an Integer Program (IP) and solving it using dynamic programming. In contrast to existing solutions where heuristics are used for scaling decision-making that results in sub-optimal solutions, we use dynamic programming to solve the IPs 
% that
which generates optimal solutions by breaking the problem into simpler subproblems and solving each subproblem only once, storing their solutions to avoid redundant calculations. For the transitions between vertical and horizontal scaling strategies, \namex{} analyzes why an autoscaler needs to switch between scaling strategies, how to transit between them, and when to do it. For that, \namex{} designs a state transition policy to decide when to use which autoscaling strategy and uses it during runtime to enable responsiveness and 
% cost-efficiency 
resource efficiency while guaranteeing end-to-end inference latency.

% \namex{} gets the application pipeline with its SLO, profiles the processing latency of the DL models in the pipeline, and calculates the queuing delay based on the given batch size and the arrival rate. Furthermore, it creates one Integer Program (IP) for each scaling mechanism and invokes them when there is a change in the workload or when the workload becomes stable again. In contrast to existing solutions where heuristics are used for scaling decision-making that results in sub-optimal solutions, we use dynamic programming to solve the IPs that generate optimal solutions by breaking the problem into simpler subproblems and solving each subproblem only once, storing their solutions to avoid redundant calculations.

In short, this paper makes the following contributions. 
After presenting the motivation and identifying the challenges (\S\ref{sec:motivation}), we 
\begin{itemize}
    \item present the design of \namex{}, including its overall architecture and system components (\S\ref{sec:design});
    \item introduces an integer program to capture the vertical and horizontal resource scaling problem in DL inference serving systems and present our solutions based on dynamic programming (\S\ref{sec:solution});
    \item build a system prototype for \namex{} and evaluate it with real-world workload traces (\S\ref{sec:evaluation}). 
    Overall, \namex{} reduces the SLO violation by over $10\times$ compared to the baselines.
\end{itemize}
\S\ref{sec:relatedwork} summarizes related work. \S\ref{sec:conclusion} draws final conclusions.

%% file: sections/sec_2_motivation.tex
\section{Background and Motivation} \label{sec:motivation}

Inference services, a critical component of online platforms and mainly composed of multiple models that are chained together, have unique characteristics that make them both latency-sensitive and resource-intensive. The latency sensitivity of these services arises from their interaction with online users. Users expect quick responses when interacting with these services, making delay or latency undesirable. On the other hand, inference services are also resource-intensive due to the heavy computations they need to perform~\cite{2020-osdi-serving, 2020-socc-inferline}.
The dual characteristics of latency sensitivity and resource-intensiveness make designing and managing inference services challenging. It is essential to find a balance between providing quick responses to user interactions and managing the heavy computational resources required by these services.
\begin{figure}[!t]
    \centering
    \includegraphics[width=0.48\textwidth]{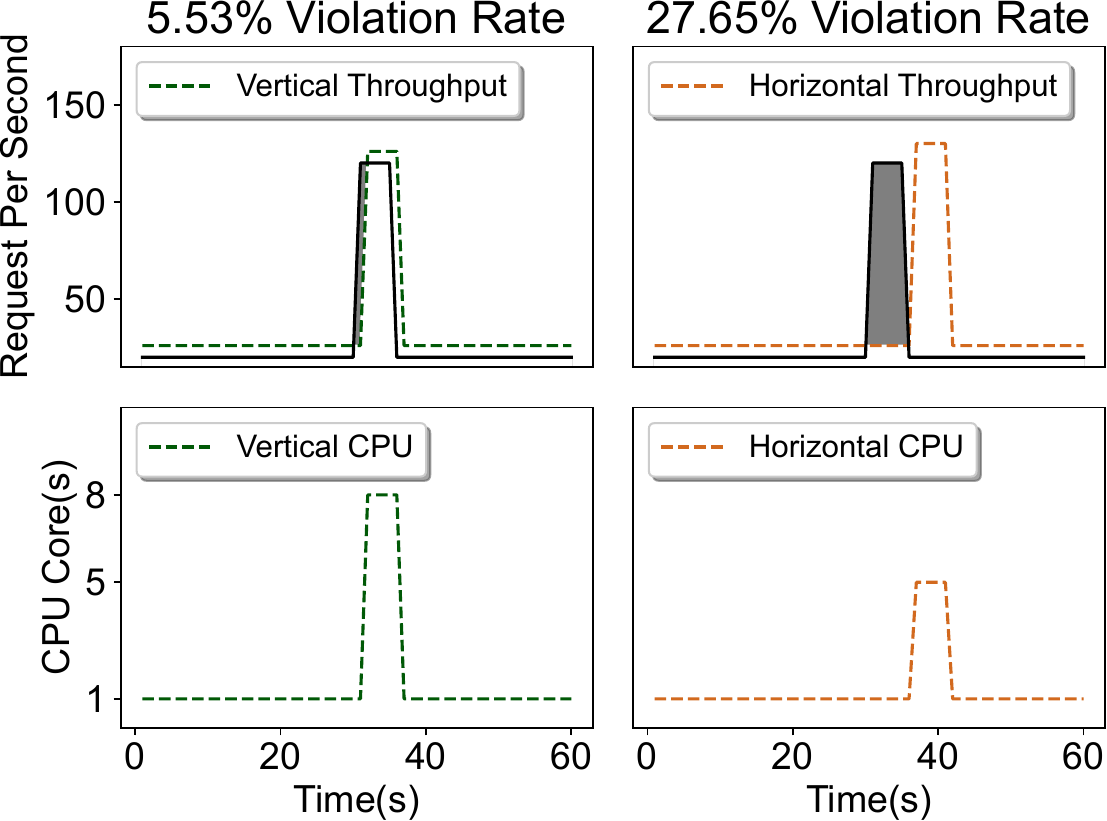}
    % \vspace{-0.2cm}
    \caption{Vertical scaling vs. horizontal scaling reaction time in case of workload bursts. Horizontal scaling is not responsive. The gray area indicates SLO violation.}\label{fig:vertical_horizontal_short}
    \vspace{-0.2cm}
\end{figure}
\begin{figure}[!t]
    \centering
    \includegraphics[width=0.48\textwidth]{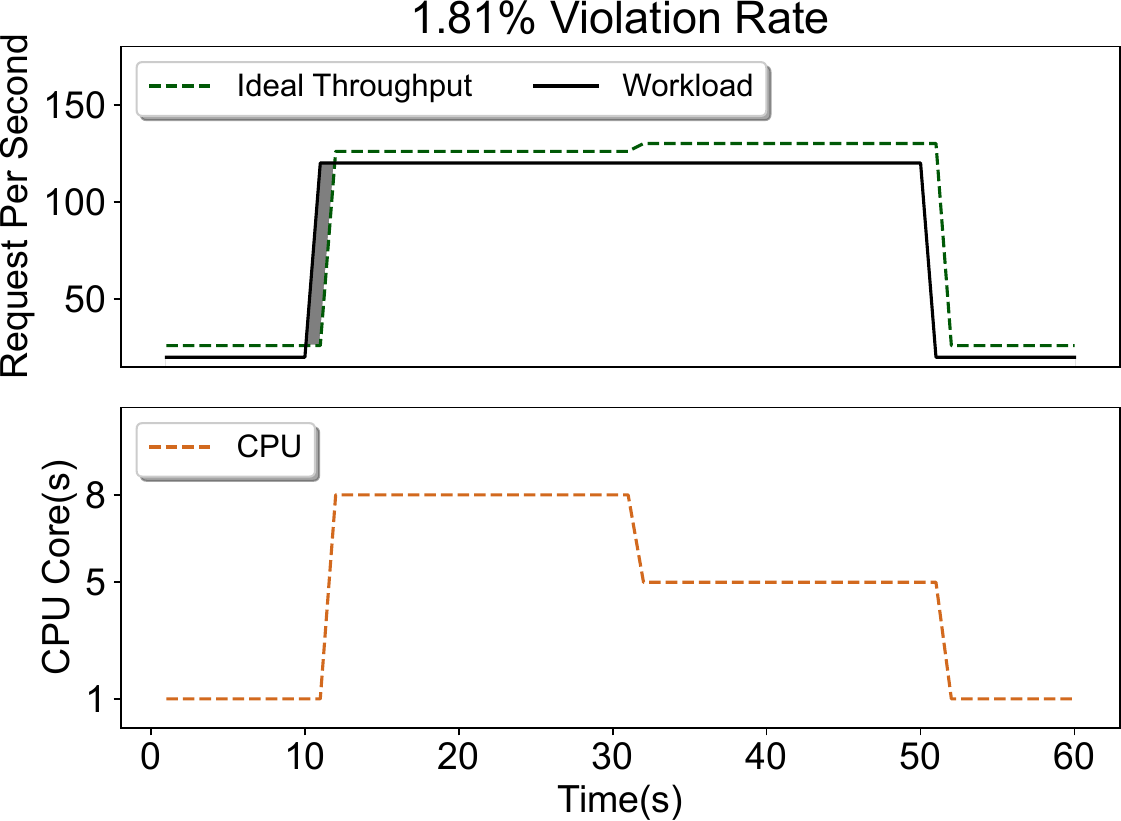}
    % \vspace{-0.2cm}
    \caption{Using vertical and horizontal scaling jointly to absorb bursts and reduce operational costs. Vertical scaling provides responsiveness, while horizontal scaling provides cost efficiency. The gray area indicates SLO violation.}\label{fig:vertical_horizontal_long}
    \vspace{-0.6cm}
\end{figure}

Figure~\ref{fig:vertical_horizontal_short} depicts the vertical and horizontal scaling response time of a ResNet car detection model~\cite{he2016deep} to the workload changes under the SLO of 1000~ms. The number of requests starts from 20 requests per second (RPS) and then rises $6\times$ (to 120~RPS) for 5 seconds and then drops back to 20 RPS. If the system captures the workload changes immediately and uses horizontal scaling, the new instances will become available after 5-6 seconds as evaluated in~\cite{razavi2022fa2}. In this scenario, all the requests (100) in that short period will violate their SLO, and the scaling decisions cannot capture the burstiness and waste resources until a new scaling decision is made. On the other hand, by using in-place vertical scaling, we can change the computational resources of the available DL models and absorb the burstiness almost instantly (with an overhead of less than 100~ms). 
Next, as the workload decreases, we reduce the allocated resources, reducing operational costs. Using in-place vertical scaling for a short duration of 5 seconds, we could reduce the SLO violation rate by roughly $5\times$. It should be noted that the amount of resources in vertical scaling is higher than in horizontal scaling, as discussed in the previous section, meaning that horizontal scaling provides higher throughput. The next question arises: can we switch to horizontal scaling after we absorb the burstiness with vertical scaling? Figure~\ref{fig:vertical_horizontal_long} answers this question by suggesting that when the workload becomes stable again (the stabilization is discussed in Section~\ref{sec:transition:vth:how}), we can switch to horizontal scaling and reduce the operational cost by 40\% in this simple car detection scenario. Therefore, a combination of vertical and horizontal scaling can react to workload changes fast enough, simultaneously reducing total resource consumption.

\begin{table}[!t]
\caption{Comparison of \namex{} with previous works; Pipeline: A chain of models or just a single model? Resource Efficiency: Does this work use horizontal scaling for the highest resource efficiency? Responsiveness: Does this approach use in-place vertical scaling to respond quickly to the changes in the workload? ($\ast$) uses model-variants as a way of vertical scaling.}
\label{table:comparison}
\centering
\small
\begin{tabular}{@{} l c c c @{}}
\toprule
\textbf{System} & \textbf{Pipeline} & \textbf{Resource Efficiency} & \textbf{Responsiveness} \\
\midrule
INFaaS & \textcolor{red}{\ding{53}} & \textcolor{teal}{\ding{51}} & \textcolor{red}{\ding{53}}$\ast$ \\
InferLine & \textcolor{teal}{\ding{51}} & \textcolor{teal}{\ding{51}} & \textcolor{red}{\ding{53}} \\
GrandSLAm & \textcolor{teal}{\ding{51}} & \textcolor{red}{\ding{53}} & \textcolor{red}{\ding{53}} \\
FA2 & \textcolor{teal}{\ding{51}} & \textcolor{teal}{\ding{51}} & \textcolor{red}{\ding{53}} \\
Scrooge & \textcolor{teal}{\ding{51}} & \textcolor{teal}{\ding{51}} & \textcolor{red}{\ding{53}} \\
Cocktail & \textcolor{red}{\ding{53}} & \textcolor{red}{\ding{53}} & \textcolor{teal}{\ding{51}} \\
InfAdapter & \textcolor{red}{\ding{53}} & \textcolor{teal}{\ding{51}} & \textcolor{red}{\ding{53}}$\ast$ \\
IPA & \textcolor{teal}{\ding{51}} & \textcolor{teal}{\ding{51}} & \textcolor{red}{\ding{53}}$\ast$ \\
Sponge & \textcolor{red}{\ding{53}} & \textcolor{red}{\ding{53}} & \textcolor{teal}{\ding{51}} \\
\namex{} & \textcolor{teal}{\ding{51}} & \textcolor{teal}{\ding{51}} & \textcolor{teal}{\ding{51}} \\
\bottomrule
\end{tabular}
\end{table}

Table \ref{table:comparison} presents an overview of related latency-sensitive inference serving systems. While most of the previous works use only horizontal scaling or model-variant changes for resource efficiency, they all suffer significantly from rapid changes in the workload if they do not over-provision their resources, which directly impacts operational costs negatively. On the other hand, the only work that considers responsiveness using vertical scaling in terms of changing the resources (not the model variant or model ensemble) becomes unpractical when the workload surpasses the capacity of one DL model with the highest possible resource allocation (hardware limitation). Furthermore, it does not consider pipeline, hence the challenges of data dependency using in-place vertical scaling are not addressed.

%% file: sections/sec_3_system_design.tex
\section{\namex{}} \label{sec:design}

In this section, we first discuss the challenges of designing a new autoscaler that benefits from both vertical and horizontal autoscaling mechanisms. 
Next, we give an overview of \namex{}, its architecture, and the main components.
Finally, we discuss the assumptions and the limitations of \namex{}.

\subsection{Challenges} \label{sec:challenges}
We have identified the following challenges in designing a new strategy that symbiotically combines both horizontal and vertical scaling:

\noindent\textit{Data dependency \textnormal{\textsc{[dd]}}.} The resource allocation decision becomes even more challenging when DL models have data dependencies (pipelines) in the serving system since the resource scaling decision for one DL model may affect downstream DL models. Therefore, we need a fast enough approach to find solutions for all the DL models in the pipeline in a single shot.

\noindent\textit{Adaptation period \textnormal{\textsc{[ap]}}.} After using vertical scaling, we need to identify when is a good time to switch to horizontal scaling to save resources. To be precise, we need to find an answer to this question: ``How to correctly predict that the workload is stable enough that we do not need to allocate extra resources for vertical scaling?''

\noindent\textit{Hardware limitation for vertical scaling \textnormal{\textsc{[hl]}}.} One of the drawbacks of vertical scaling compared to horizontal scaling is that the DL model's maximum capacity is bound to the hardware capabilities. Every hardware has a maximum limit for processing power; once it is reached, the DL model cannot scale up any further. Thus, we must find a placement algorithm for assigning DL models in a pipeline in a distributed cluster.

\noindent\textit{Batch size \textnormal{\textsc{[bs]}}.} Apart from resource allocation and resource placement in inference serving systems, there is a dominant factor that directly affects the latency and throughput of DL models, named batch size. The batch size can be changed online to increase the DL model's throughput based on the demand at the cost of higher inference latency. As a result, we need to find the optimal batch size. Large batches can significantly impact the response time of requests within a batch. On the other hand, small batches might miss out on opportunities for improved throughput and cost efficiency.

\subsection{System Design}
\begin{figure}
    \centering
    \includegraphics[width=0.49\textwidth]{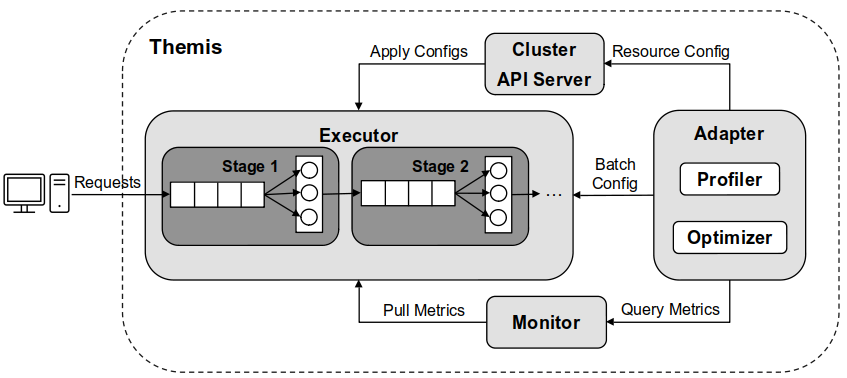} %{figures/empty_box.png}
    % \vspace{-0.1cm}
    \caption{An overview of the \namex{} architecture. The executor receives the requests and processes them. The monitor service collects the metric data from the stages. The optimizer makes both vertical and horizontal scaling decisions for the DL models. The Adapter enforces the scaling decisions by configuring the queues and the executor.}\label{fig:system_design}
    \vspace{-0.4cm}
\end{figure}

Figure ~\ref{fig:system_design} provides an overview of \namex{}. The system comprises five main parts (profiler, executor, monitor, optimizer, and adapter). The profiler creates a new performance model for any registered DL model in the system. The stage component gets the requests and executes the DL model. The monitor component keeps track of the pipeline's request rate (workload distribution). The optimizer uses data from the monitor and the profiler components to find an optimal resource allocation solution based on in-place vertical or horizontal scaling. Finally, the adapter enforces the decisions made by the optimizer to the system.

The \textbf{profiler} component sends profiling requests to the DL model with different configurations (different CPU and batch size allocations) (answering \textnormal{\textsc{[bs]}}). It then records the latencies under each configuration and uses them to create the performance model of each DL model in the system. Furthermore, these latency profiles are used for in-place vertical scaling in the optimizer. Note that this component runs offline, and the performance profiles generated by this component will be used online.

The \textbf{executor} consists of multiple stages, each of which is composed of the queuing and the model processing parts. Each stage has one centralized queue and one or more processing model instances. The queuing component fetches the requests from the user or other stages, sends the arrival rate statistic to the optimizer, batches the requests based on the optimizer's decision, and sends the batches to their associated models for inference. It uses a round-robin approach to distribute batches among the processing instances (decided by the optimizer) in its stage. The processing instance has the DL model and the computing power (determined by the optimizer based on the incoming workload) to process the batches sent by the stage queuing component. After processing the batches, the instance sends the results to the next stage or the user. Both components keep track of the queuing and processing latencies. In the last stage, a query with the gathered latencies is generated and sent to the monitor component for further analysis.

The \textbf{monitor} gets the arrival rate statistics from the first stage queue and the total request latency from the last model instances. The optimizer will fetch the arrival rate statistics to find optimal resources and batch sizes for the whole system while guaranteeing end-to-end request latencies, and the total request latency will be used to report the system and the profiler's performance. To reduce the monitor overhead, we append the latencies to the same request as they pass through different pipeline stages.

The \textbf{optimizer} gets the arrival rate from the first queuing component and feeds it to the horizontal autoscaler. Next, it uses an LSTM predictor to estimate the highest workload in the following 10-second time window and feeds the predicted workload to the horizontal autoscaler (answering \textnormal{\textsc{[ap]}}). The horizontal scaling results are sent to the adapter if no additional resources are required. If not, the optimizer checks if there are enough physical hardware resources for using in-place vertical scaling to support the current workload (answering \textnormal{\textsc{[hl]}}). If there are enough resources, the system uses in-place vertical scaling to respond to changes in the workload by finding the optimal resource allocation for all the DL models in the system in a single shot (answering \textnormal{\textsc{[dd]}}). If not, the optimizer calculates the maximum throughput using vertical scaling and switches to horizontal scaling for the remaining requests.

The \textbf{adapter} is responsible for enforcing the configurations made by the optimizer to the cluster. The configurations for each DL model in the system are batch size, resource allocation of model instances, and the number of instances. After the new configuration is received from the optimizer, the adapter first compares the new set of configurations with the current state of configurations. If there are changes, the adapter makes two calls for each model's new configuration, one to the queue for adjusting the batch size and one to the executor to adjust the computing resources of the models (vertical scaling) or change the instance number of the model (horizontal scaling). The batch size will be immediately updated, and changes in the computing resources will be almost instant (less than 100~ms). The scale in/out of the instances will take seconds (roughly 5-6 seconds).

\subsection{Assumptions and Limitations} \label{sec:limitations}
The in-place vertical scaling part of \namex{} uses the vertical scaling feature of Kubernetes, which currently is on the alpha branch and only supports CPU and memory real-time modification. However, inference serving systems can benefit greatly from other accelerators, such as GPUs and TPUs, which, unfortunately, are not supported at this moment. 
Also, \namex{} uses simple queuing management, whereas a more complex queuing simple may provide a more accurate queuing latency. 
Moreover, multiple variants for the same DL models with different properties, such as accuracy/latency trade-offs, can be leveraged to reduce costs further. 
Finally, we consider uniform SLOs, where all DL inference requests in the system have the same end-to-end latency SLO requirement.

%% file: sections/sec_4_problem_statement.tex
\section{Autoscaling Problem} \label{sec:solution}
In this section, we provide a mathematical representation of the joint horizontal-vertical autoscaling problem in an inference serving system to react to the changes in the workload. We encapsulate the joint autoscaling problem into an Integer Program (IP) that decides the batch size, computing resource, and the number of instances of each DL model in the system in a single shot. The goal of the IP is to minimize the cost while guaranteeing the requests SLO. It achieves it by using vertical scaling when there are sudden changes in the workload to capture the workload burstiness. Furthermore, when the workload is stable (see Section~\ref{sec:transition:vth:how}), the IP switches to horizontal scaling to save operational costs. We use different dynamic programming approaches to solve the IP based on the workload status.

%in case of a sudden increase in the workload (vertical scaling) and reduce the total cost. In contrast, the workload is stable (horizontal scaling). We further use different dynamic programming approaches to solve the IP based on the workload status.

\subsection{Performance Profiling} \label{sec:performance_model}
A DL model processes requests with different latencies based on the allocated computational resources and batch sizes. Previous works have shown that processing latency has a linear/quadratic relationship with batch size~\cite{2019-eurosys-grandslam, razavi2022fa2}. Sponge~\cite{razavi2024sponge} has shown that the processing latency relationship is inverse to the core allocation in inference serving systems. We follow the same guidelines and use the following formula to build a performance profile for each DL model in the system:

\vspace{-5pt}
\begin{equation}
    \begin{split}
    \label{eq:latency_batch_core}
        l(b, c) = & \frac{\gamma \times b}{c} + \frac{\epsilon}{c} + \delta \times b + \eta.
    \end{split}
\end{equation}

Formula~\ref{eq:latency_batch_core} predicts the processing latency with the given batch size ($b$) and computing resources ($c$) of the DL model. We execute the DL model with different batch sizes and resource allocation configurations to find the static variables ($\gamma, \epsilon, \delta, \eta$) to get the processing latencies. We then use the latency data and fit the above formulation to get the DL performance profile. We perform the same approach offline for all the DL models in the system and create a unique performance profile for each.

All the used notations are available in Table~\ref{table:notation}.
\begin{table}[!t]
    \centering
    \small
    \caption{Notations}\label{table:notation}
    \begin{tabular}{ @{}p{1.2cm}p{5.8cm}@{}  }
        \toprule
        \textbf{Symbol} & \textbf{Description}\\
        \midrule
        $S$ & Set of all models for an application \\
        $SLO$ & Latency SLO of the application \\
        $n_s$ & number of instances of model $s \in S$\\
        $b_s$ & batch size of model $s \in S$\\
        $c_s$ & CPU core allocation of model $s \in S$\\
        $\lambda$ & Request arrival rate \\
        $l_s(b, c)$ & Processing time of model $s \in S$ with allocation core $c$ and batch size $b$ \\
        $q_s(b)$ & Queuing time of model $s \in S$ with batch size $b$\\
        %$q_s(b, c, n)$ & Queuing time of model $s \in S$ with allocation core $c$, batch size $b$, and $n$ processing instances\\
        $h_s(b, c)$ & Throughput of model $s \in S$ with allocation core $c$ and batch size $b$ \\
        \bottomrule
    \end{tabular}
    % \vspace{-0.4cm}
\end{table}

\subsection{Queue} \label{sec:queue}
Due to the differences in the number of DL instances, computing resources, and batch size, the DL model instances process requests with different latencies. Also, to form the batches to increase the system performance, the first request in the batch must wait for the arrival of the last request in the batch, causing an extra queuing delay. Because of the mentioned reasons, the requests in batches leave the queue at different speeds (i.e., the last request in the batch leaves the queue immediately, and the first request in the batch must wait the longest, hence it has the slowest speed in terms of leaving the queue), causing a dynamic queuing latency. 
FA2~\cite{razavi2022fa2} uses Equation~(\ref{eq:queue_fa2}) for the worst-case scenarios. The worst-case scenario can happen when there are no available instances (out of the n instances for the same DL model) to process the current batch (meaning that all $n$ instances are busy and the next batch needs to wait for one of the instances to be free ($l(b)-\frac{nb+1}{\lambda}$)) or the first request waits for the arrival of the last request in the same batch ($\frac{b-1}{\lambda}$). Therefore, the worst case is the maximum of these two equations, as shown in the following equation.
\begin{equation} \label{eq:queue_fa2}
    q(b,n) = \max \left(\frac{b-1}{\lambda}, l(b)-\frac{nb+1}{\lambda}\right).
\end{equation}

However, they do not consider the speed-up/down caused by changing the computing resources of the DL model instances, which directly affects the queuing drainage. Therefore, we propose the following equation to incorporate the computing resources in the DL model as well:

\begin{equation} \label{eq:queue_with_cpu}
    q(b,c,n) = \max \left(\frac{b-1}{\lambda}, l(b,c)-\frac{nb+1}{\lambda}\right).
\end{equation}

Similar to FA2, we argue that the worst-case scenario happens for the latter scenario, where the first request needs to wait for the last request of the same batch due to the fact that by increasing the computing resources and reducing the processing latency, the DL models catch up with the workload. This means that the aggregate throughput of all the instances will equal or exceed the workload, i.e., $n \times h(b,c) \ge \lambda$, i.e., the second part of the above equation is always less than zero. Therefore, we use a simplified queuing latency prediction as: 

\begin{equation} \label{eq:queue}
    q(b) = \frac{b-1}{\lambda}.
\end{equation}

\subsection{Problem Formulation}
The goal of the optimizer is to decide the least amount of allocated resources of all instances ($\sum_{s \in S} n_s \times c_s$), subject to guaranteeing the end-to-end latency of requests and supporting the workload.

The end-to-end latency of a request is the aggregation of queuing and processing latencies of all the stages in the pipeline, which should be lower than the given SLO:
\begin{equation} \label{eq:total_latency}
    \text{End-to-End Latency} = \sum_{s \in S} l_s(b_s, c_s) + q_s(b_s) \le SLO.
\end{equation}

To maintain system stability, the combined throughput of all instances of a stage must meet or exceed the request rate. Formally, for any stage, the product of its instance's throughput and the number of instances should be greater than or equal to the sum of request rates across all instances, i.e., \( s \in S \), \( h_s(b_s, c_s) \times n_s \geq \lambda \). This constraint ensures adequate provisioning for all stages, effectively managing the queuing of inference requests at each stage.

The problem can be formulated with the following IP:
\begin{equation}
    \begin{aligned}
        \min &&& \sum_{s \in S} n_s \times c_s \\
        \text{subject to} &&& \sum_{s \in S} l_s(b_s, c_s) + q_s(b_s) \le SLO\\
        &&& \lambda \le h_s(b_s, c_s) \times n_s , \forall s \in S\\
        &&& b_s, c_s, n_s \in \mathbb{Z}^+, \forall s \in S
    \end{aligned}
    \label{eq:ip}
\end{equation}

% In the optimization objective, we added a minor penalty for the batch sizes to use the minimum batch per stage for two reasons: first, increasing batch sizes cannot reduce the number of instances anymore, resulting in a higher processing latency per request, and second, the queuing delay will be increased since the first request in the batch should wait for the last request. Both reasons increase the total latency, which may violate the SLO unnecessarily due to the performance profiling error. 
The last constraint in Equation~\ref{eq:ip} states that the batch sizes and number of instances should be positive. 
Next, we solve the above IP with different approaches based on the current workload situation.

\subsection{Optimizer}
Two dynamic variables per stage affect the total cost significantly and can be adjusted in the IP from the previous section, namely core allocation, $c$, and instance number, $n$. The former is used in vertical scaling (changing the processing latency of a model), and the latter is used in horizontal scaling (distributing the workload to multiple instances). As we explain in Section~\ref{sec:transition}, vertical scaling is not the most cost-efficient mechanism for serving requests compared to horizontal scaling, but the in-place vertical scaling feature is more responsive when there are changes in the workload. Therefore, we solve the IP with two different mindsets, using vertical scaling to respond quickly or saving more resources with horizontal scaling. 

However, a similar IP cannot be solved efficiently in real-time as the problem grows exponentially as the number of stages grows~\cite{romero2019infaas, razavi2022fa2, ghafouri2024ipa}. Consequently, we need to use heuristics or limit the solver to some extent (reducing the feasible space to explore). We leverage the limited SLO (which is usually a few thousand milliseconds), limited batch sizes (1-16), and limited core allocation (limited to the hardware the model is placed on (1-16 CPU cores)) in inference serving systems and provide dynamic programming algorithms to find the optimal solution for either vertical or horizontal scaling. Both algorithms run in $O(SLO \times b_{max} \times c_{max} \times |S|)$ time complexity, which reduces the exponential execution time (dependent on the number of stages) to the dominant SLO time by incorporating the space complexity using dynamic programming. Section~\ref{sec:transition:vth:when} discusses which algorithm to be executed at a given time.

\subsubsection{Vertical Scaling}
As discussed above, we use dynamic programming to find the optimal solution for the vertical scaling to absorb the sudden changes in the workload. 

Algorithm~\ref{algo:dp_vertical} aims to find the optimal CPU allocations for the given pipeline considering their performance profile and the specified SLO. The algorithm takes the pipeline SLO, a set of models denoted by $S$ with their corresponding performance profile, and the workload $\lambda$ as input and returns the values $c_s$ and $b_s$ for all the DL models $s \in S$. Dynamic programming solves the IP by first solving the problem with just having one stage and then increasing it to the pipeline one by one.
For any added model, we consider all possible SLOs (1 to SLO) and divide the picked number into two parts, one for the current DL model and one for the previous DL models. We check whether it is possible with the current division to serve the workload in all the DL models.
We continue the procedure until all the models are visited. In the end, if there is no possible solution, we use a binary search on the workload to find the maximum possible workload supported by vertical scaling and bring up new instances (horizontal scaling) to support the remaining workloads.

\begin{algorithm}[!t]
\small
\SetAlgoLined
\SetKwInOut{Input}{input}\SetKwInOut{Output}{output}
\SetKwIF{If}{ElseIf}{Else}{if}{then}{else if}{else}{end if}%
\Input{SLO, Set of models $S$ and their latency model, $\lambda$} 
\Output{$c_s, b_s \forall s \in S$}
% dp = [|S|][False] * (SLO + 1)\\
% dp[0][0] = True\\
% best = [|S|][(1000, 1000, 1000)] * (SLO + 1)\\
% best[0][0] = (0, 0, 0)\\
\For{$s$ in $[0, |S|, 1]$}{
    \For{$i$ in $[SLO, -1, -1]$}{
        \uIf{ (s == 0 and dp[s][i]) or s (!= 0 and dp[s - 1][i])} {
            \For{$c$ in $[1,c_{max}]$}{
                \For{$b$ in $[1,b_{max}]$}{
                    l = calculate $l_s(b, c)$, q = $\frac{b - 1}{\lambda}$\\
                    h = calculate $h_s(b, c)$ \\
                    l += q\\
                    \uIf{$h \le \lambda$ and $l \le SLO$} {
                        \textbf{continue}
                    }
                    \uIf{s == 0} {
                        \uIf{$dp[s][i + l]$ is False} {
                            $dp[s][i + l]$ = True\\
                            $best[s][i + l]$ = $(c, c, b)$
                        }
                    }
                    \uElseIf{$dp[s - 1][i]$  and i > 0} {
                        \uIf{$dp[s][i + l]$ is False} {
                            $dp[s][i + l]$ = True\\
                            $best[s][i + l]$ = $(best[s - 1][i][0] + c, c, b)$
                        }
                        \uElseIf{$best[s - 1][i][0] + c < best[s][i + l][0]$} {
                            $best[s][i + l]$ = $(best[s - 1][i][0] + c, c, b)$
                        }
                    }
                }
            }
        }
    }
}
Check if there is a solution in \textit{dp[|S| - 1]}.\\
% $least_c$ = 1000\\
% ind = -1\\
% \For{$i$ in $[1, SLO + 1, 1]$}{
%     \uIf{$d[|S| - 1][i]$} {
%         \uIf{$ind == -1$}{
%             ind = i//
%             $least_c$ = $best[|S| - 1][i][0]$
%         }
%         \uElseIf{$best[|S| - 1][i][0] < least_c$} {
%             ind = i\\
%             $least_c$ = $best[|S| - 1][i][0]$
%         }
%     }
% }
% \uIf{$ind == -1$}{
\uIf{No solution}{
        left = 1, right = $\lambda$\\
        \While{right - left > 1} {
            mid = (right - left) // 2\\
            \uIf{Vertical scaling with $\lambda = mid$} {
                right = mid
            }
            \Else{
                left = mid
            }
        }
        \textbf{return} Vertical Scaling with $\lambda = left$ and Horizontal Scaling with $\lambda = \lambda - left$ with the same CPU as Vertical Scaling.
}
\Else {
    % result = [], c = |S| - 1\\
    % ind = index of the found solution in the dp[|S| - 1]\\
    % \While{$ind != 0$} {
    %     result.append((best[c][ind][1], best[c][ind][2])\\
    %     ind -= $l_s(best[ind][1], best[ind][2])$
    %     }
    result = calculate recursively the optimal configurations for all the models using \textit{best}.\\
    \textbf{return} result
}
\caption{Vertical Scaling}
\label{algo:dp_vertical}
\end{algorithm}

To be precise, we iterate first over the stages (line 1) and then iterate on all possible SLO values with different batch sizes and core allocations (lines 2-5). Then, we estimate the total latency by calculating the processing latency and then aggregating it with the queuing latency (lines 6-8).
Next, if the current throughput supports the incoming workload and the aggregated latency is lower than the SLO, we check if the current model is the first in the pipeline since there is no need to divide the current SLO. If so, we store the current configuration (current CPU and batch size allocations) (lines 11-14). If not, we check if the current configuration is the least total resource allocations with the current SLO division (lines 15-20).
After considering all the configurations, we check if there is a possible candidate (line 21) by checking if the algorithm has reached the last DL model in the pipeline. If not, we use binary search on the workload and feed it to the same algorithm to find how much of the workload vertical scaling can support (lines 22-29). We then calculate the needed instances for serving the remaining requests using horizontal scaling with the same CPU core allocations (line 30). If there is a valid configuration, we find the configurations for all the models recursively (lines 31-33).

\subsubsection{Horizontal Scaling}
Similar to vertical scaling, we use dynamic programming to find the optimal instance number (horizontal scaling) for all the DL models in the system in a single shot. The main difference between the vertical and horizontal scaling algorithms is that we can support any workload using horizontal scaling. Hence, we calculate the number of needed 1-core instances (line 5), and we calculate the least total resource-consuming configuration (lines 11-20) in Algorithm~\ref{algo:dp_horizontal}. Finally, we recursively find the optimal number of instances for all the models and send the results to the adapter for the system adaptation.

\begin{algorithm}[!t]
\small
\SetAlgoLined
\SetKwInOut{Input}{input}\SetKwInOut{Output}{output}
\SetKwIF{If}{ElseIf}{Else}{if}{then}{else if}{else}{end if}%
\Input{SLO, Set of models $S$ and their latency model, $\lambda$} 
\Output{$n_s, b_s \forall s \in S$}
% dp = [|S|][False] * (SLO + 1)\\
% dp[0][0] = True\\
% best = [|S|][(1000, 1000, 1000)] * (SLO + 1)\\
% best[0][0] = (0, 0, 0)\\
\For{$s$ in $[0, |S|, 1]$}{
    \For{$i$ in $[SLO, -1, -1]$}{
        \uIf{ (s == 0 and dp[s][i]) or s (!= 0 and dp[s - 1][i])} {
            \For{$b$ in $[1,b_{max}]$}{
                l = calculate $l_s(b, 1)$, q = $\frac{b - 1}{\lambda}$\\
                h = calculate $h_s(b, 1)$ \\
                l += q\\
                \uIf{$h \le \lambda$ and $l \le SLO$} {
                    \textbf{continue}
                }
                ins = workload // throughput\\
                \uIf{s == 0} {
                    \uIf{$dp[s][i + l]$ is False} {
                        $dp[s][i + l]$ = True\\
                        $best[s][i + l]$ = $(ins, ins, b)$
                    }
                }
                \uElseIf{$dp[s - 1][i]$  and i > 0} {
                    \uIf{$dp[s][i + l]$ is False} {
                        $dp[s][i + l]$ = True\\
                        $best[s][i + l]$ = $(best[s - 1][i][0] + ins, ins, b)$
                    }
                    \uElseIf{$best[s - 1][i][0] + ins < best[s][i + l][0]$} {
                        $best[s][i + l]$ = $(best[s - 1][i][0] + ins, ins, b)$
                    }
                }
            }
        }
    }
}
\caption{Horizontal Scaling}
\label{algo:dp_horizontal}
\end{algorithm}

%% file: sections/sec_5_transition.tex
\section{Transition} \label{sec:transition}
In this section, we explain the necessity of transitioning to horizontal scaling after initially using in-place vertical scaling to respond to changes in workload. We detail the circumstances under which vertical and horizontal scaling transitions should occur and provide a comprehensive method for switching between these mechanisms.

\begin{figure}
    \centering
    \includegraphics[width=0.49\textwidth]{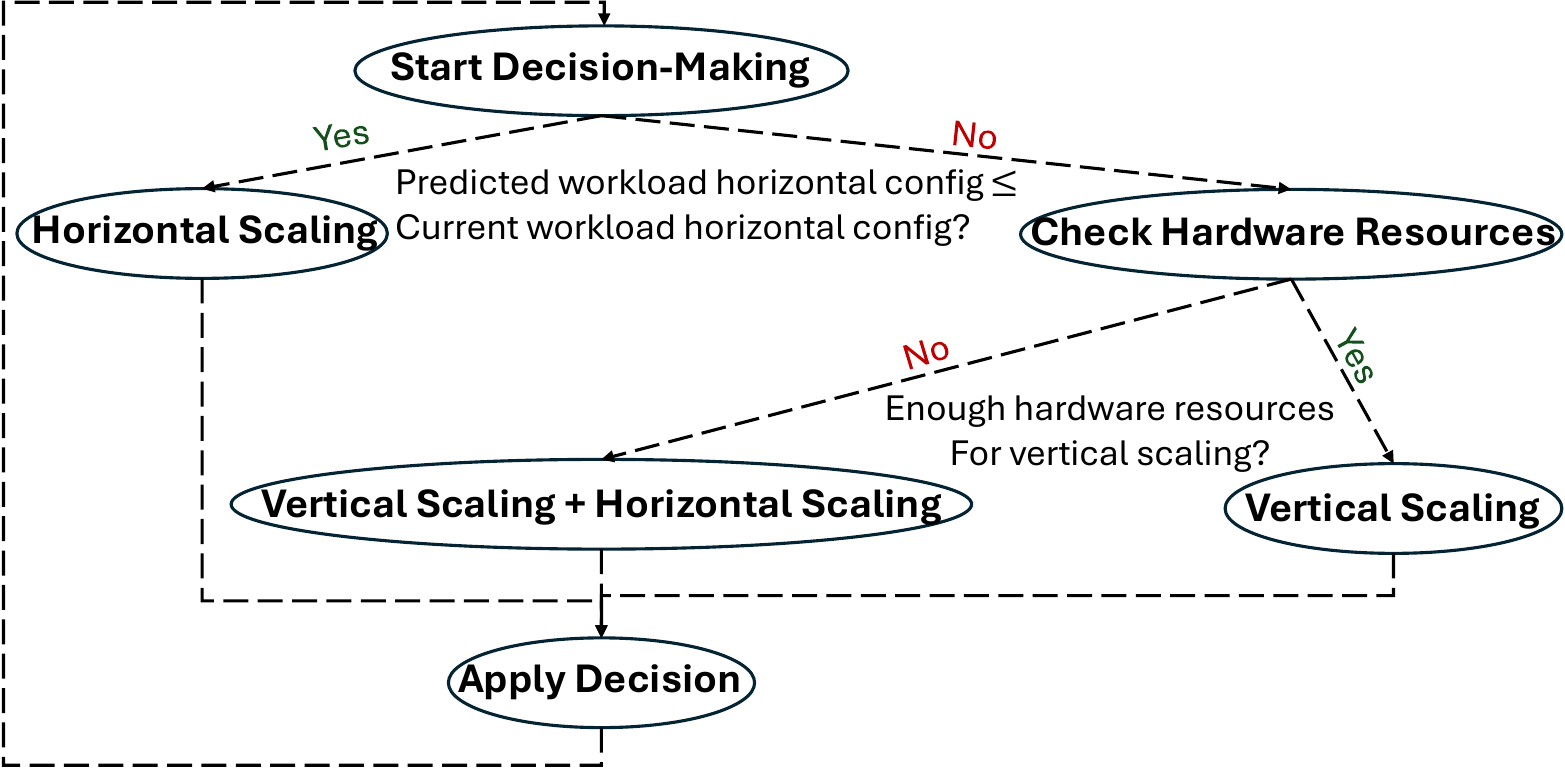}
    % \vspace{-0.1cm}
    \caption{Transition states between vertical and horizontal scaling strategies.}\label{fig:transition}
    \vspace{-0.4cm}
\end{figure}

First, we present a mathematical proof demonstrating the superior performance of horizontal scaling in terms of throughput, addressing the \textit{why} of this transition, contrasting it with the responsiveness offered by vertical scaling as discussed in Section~\ref{sec:motivation}. We outline the horizontal and vertical scaling transition steps based on this proof (answering \textit{how}). Finally, to address the \textit{when} of these transitions, we develop an LSTM model to predict the maximum requests per second (RPS) for the next few seconds, guiding the decision of when to switch from vertical scaling to horizontal scaling.
All the transition states are available in Figure~\ref{fig:transition}.

\subsection{Vertical Scaling to Horizontal Scaling} \label{sec:transition:v2g}
This part describes why we need to switch to horizontal scaling from vertical scaling and discusses how and when it must be done.

\subsubsection{Why} \label{sec:transition:vth:why}
Amdahl's law describes the theoretical speed-up in the execution time of a parallelizable task~\cite{amdahl1967validity}.
%under a fixed workload. 
%It uses the parallelization share of the task and the available resources for the task.
In Formula~\ref{eq:Amdahl}, $L(r)$ calculates the speed-up of a task with the given computational resources ($r$) and the parallelization share ($p$) of the task.

\begin{equation}
    L(r) = \frac{1}{(1 - p) + \frac{p}{r}}.
    \label{eq:Amdahl}
\end{equation}

We use Amdahl's law to show that 1-core instances always provide the same or a higher throughput compared to the multiple-core instances in DL models under a fixed workload using the same amount of resources.
We use induction to prove this statement by setting total resources $r = 2$ and then $r = n \times (n + 1)$.

For $r = 2$, we have $2 \times$ 1-core instances which implies $2 \times L(1)$ and $1 \times$ 2-core instance which gives the task speed-up of $1 \times L(2)$.
The execution time of $2 \times L(1)$ instances is $2\times(\frac{1}{(1 - p) + \frac{p}{1}}) = 2\times1 = 2$ while the latency of the execution time of $1 \times L(2)$ instances is $\frac{1}{(1 - p) + \frac{p}{2}} = \frac{2}{2 - p}$ which can be at most $2$ as the parallelization share can be between 0 and 1.
Therefore, $2 \times L(1) \ge 1 \times L(2)$, meaning that $2\times1-core$ has a total higher speed-up $1\times$ 2-core, resulting in the same or higher throughput.

For  $r = n \times (n + 1)$, we calculate $(n + 1)$ $n$-core instances ($(n + 1) \times L(n)$) and $n$ $n+1$-core instances ($n \times L(n + 1)$) as follows:

\begin{equation}
    (n + 1) \times L(n) = (n + 1) \times \frac{1}{(1 - p) + \frac{p}{n}} = \frac{(n + 1) \times n}{n - np + p}.
    \label{eq:r=n}
\end{equation}

\begin{equation}
    n \times L(n + 1) = n \times \frac{1}{(1 - p) + \frac{p}{n + 1}} = \frac{(n + 1) \times n}{n - np + 1}.
    \label{eq:r=n+1}
\end{equation}

As $0 \leq p \leq 1$, $(n + 1) \times L(n)$ process more requests than $n \times L(n + 1)$ based on Formula~\ref{eq:r=n} and Formula~\ref{eq:r=n+1}, $(n + 1) \times L(n)$ results in a higher throughput.

Finally, as for any number of cores, more instances with a lower core number achieve a higher throughput. We conclude that for any $r = n$, $n \times 1-core$ instances achieve the highest possible throughput under a fixed workload.

\subsubsection{How} \label{sec:transition:vth:how}
There are two scenarios in which we switch to horizontal scaling from vertical scaling: \i when the workload is stable, as described in the next section, and \ii when there are not enough hardware resources for vertical scaling to support the workload.

For the former scenario, we use the argument in the previous part and bring up as many 1-core instances as needed to serve the requests minus the currently running instances, which we reduce their resources to 1-core as soon as the rest of the instances are up and running. 
For instance, if currently, two 3-core instances are serving the requests and four 1-core instances can serve the same number of requests, we bring up two 1-core instances and reduce the 3-core instances' resources to 1-core instances using in-place resource scaling.

For the latter scenario, when there is a hardware limitation, or there is no speed-up possible because of the limited parallelism of the DL model, we serve as many requests as possible with the currently running instances by increasing their resources to the maximum to reduce the SLO violations and simultaneously bring up new 1-core instances to serve the remaining requests.

\subsubsection{When} \label{sec:transition:vth:when}
To decide when to switch from vertical scaling to horizontal scaling in case of workload stability, we design an LSTM to predict the next ten seconds' maximum RPS and feed the predicted value and the current RPS to the horizontal scaling algorithm~\ref{algo:dp_horizontal} to check whether the current and future configurations are the same. If so, we switch to horizontal scaling to save resources. 
We use the horizontal scaling algorithm to check if the workload is stable since the currently running instances may be able to serve more requests than the current workload. Therefore, more RPS may not need more resources.

\begin{figure}
    \centering
    \includegraphics[width=0.49\textwidth]{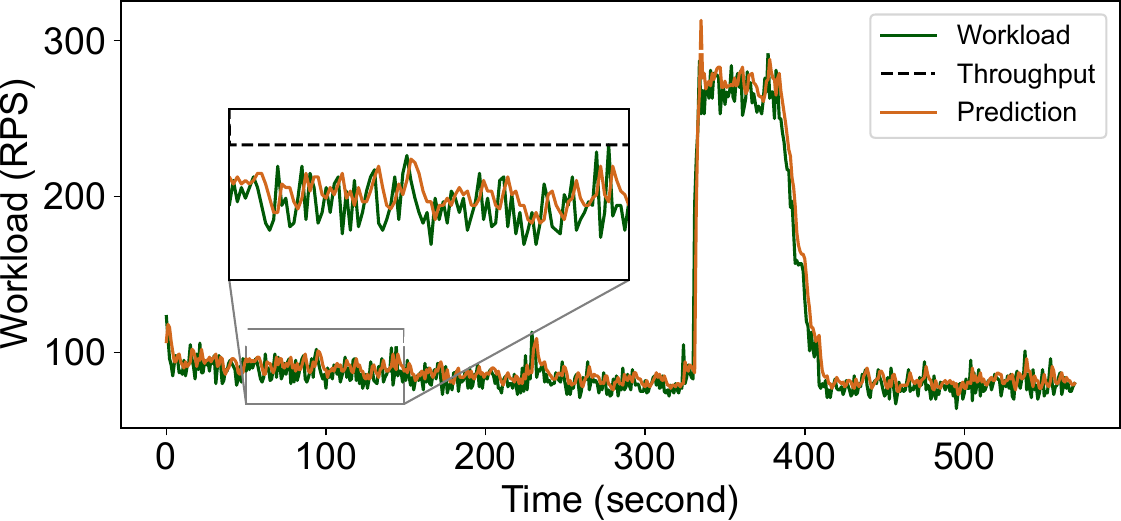}
    % \vspace{-0.1cm}
    \caption{LSTM inference result.}\label{fig:lstm}
    \vspace{-0.4cm}
\end{figure}

For building the LSTM model, we use an input layer, a 25-unit LSTM layer, followed by a one-unit dense output layer. We trained the network using the Adam optimizer with the MSE loss function.
Figure~\ref{fig:lstm} shows the LSTM result, where we train it with 14 days of the Twitter dataset~\cite{twitter-trace-2021-08} and seven days as validation (more details in Section~\ref{sec:evaluation}). The workload prediction takes less than 30~ms, with a Mean Absolute Percentage Error of 5.8\%, making it practical for getting live inference.

\subsection{Horizontal Scaling to Vertical Scaling}  \label{sec:transition:htv}
This part describes why we need to switch to vertical scaling from horizontal scaling and discusses how and when it must be done.

\subsubsection{Why and When} \label{sec:transition:htv:why}
When there is an increase in the workload, horizontal scaling needs to bring up new instances, load the required heavy libraries such as Tensorflow or PyTorch, and initialize the model. These actions take a few to tens of seconds, which drastically reduces the responsiveness of horizontal scaling. On the other hand, in-place vertical scaling can instantly increase the computing resources of the running instances, making it a better option in terms of responsiveness. Therefore, when there is a sudden surge of requests (when the current resource allocation cannot support the increased requests), we switch to vertical scaling to absorb as many extra requests as possible.

\subsubsection{How} \label{sec:transition:htv:how}
Now, to answer how to transition from horizontal scaling to vertical scaling, let's assume that we have more than one 1-core instance in a stage since with just one 1-core instance, the previous part states how to scale. The following question arises when there is a change in the workload requiring more resources: Should we increase the resources of one instance, the resources of a subset of instances, or the resources of all the instances to react to the changes in the workload? To answer this question, we use similar proof as in the previous section by first considering two 1-core instances, analyzing how to scale them vertically, and then generalizing them to any number of instances. 

For  $r = 2 \times n$, we calculate both configurations speed-up; once giving all the resources to one instance ($instance_1 = 2 \times n - 1$ and $instance_2 = 1)$ and once distribute the resources evenly ($instance_1 = instance_2 = n$). The speed-up calculation for $r = 2 \times n - 1$ is as follows:
\begin{equation}
     L(2 \times n - 1) = \frac{1}{(1 - p) + \frac{p}{2 \times n - 1}} = 1 + \frac{2 \times p \times (n - 1)}{2 \times (n - np + p) - 1}.
    \label{eq:r=2n-1,1}
\end{equation}

By having the second instance with 1-core and no speed-up, we have the total speed-up of $2 + \frac{2 \times p \times (n - 1)}{2 \times (n - np + p) - 1}$. On the other hand, if we distribute the resources evenly to both instances, each with n CPU cores, we will have the total speed-up of:
\begin{equation}
     2 \times L(n) = 2 \times \frac{1}{(1 - p) + \frac{p}{n}} = 2 + \frac{2 \times p \times (n - 1)}{n - np + p}.
    \label{eq:r=n,n}
\end{equation}

Now, if $n - np + p \le 2 \times (n - np + p) - 1$, we have shown that distributing resources evenly has a higher total speed-up, resulting in a higher throughput.
The equation can be proven as follows:
\begin{equation}
\begin{split}
    n - np + p & \le 2 \times (n - np + p) - 1\\
    1 & \le  n - np + p \\
    1 - p & \le n \times (1 - p)
\end{split}
\end{equation}
The last part is true since $n$ (the number of allocated CPU cores) is always greater than 1.

The same proof applies to any number of instances, which is omitted to reduce repeatability. Therefore, evenly distributing resources generates a higher throughput compared to giving a subset of instances more resources. We follow the same practice in our system, and when vertical scaling is needed, we scale all the instances to the same amount of resources.

%% file: sections/sec_6_experimental_results.tex
\section{Experimental Evaluation} \label{sec:evaluation}
We implemented the inference services using PyTorch~\cite{paszke2019pytorch} and Hugging Face~\cite{huggingface}. The queuing component is implemented using Python AsyncIO~\cite{python_asyncio}.
We implemented \namex{} in Python~\footnote{\href{https://github.com/mehransi/Themis}{https://github.com/mehransi/Themis}}.
We evaluated \namex{} on a testbed consisting of two servers, each equipped with Core i9-9940x CPUs at 3.3GHz and interconnected with a stable private Ethernet network (1Gbps). 
We set up a Kubernetes cluster using MicroK8s~\cite{microk8s} on these two servers and configured the InPlaceVerticalScaling feature gate, an alpha-stage feature that enables in-place vertical scaling.

\noindent\textbf{Workload}: To evaluate \namex{}, we developed a workload generator that uses the real-world Twitter traces~\cite{twitter-trace-2021-08} as the workload pattern (requests per second) and sends requests to the pipelines following a Poisson distribution to mimic the workloads on data centers. Furthermore, we use roughly 10-minute traces from the validation data discussed in Section~\ref{sec:transition:vth:when} for evaluation purposes. The selected traces show a steady workload and heavy bursts, demonstrating the systems' behavior under a realistic and dynamic workload. We scale the traces for each pipeline to match the hardware capacity we run the experiments on.

\begin{figure}
    \centering
    \includegraphics[width=0.95\linewidth]{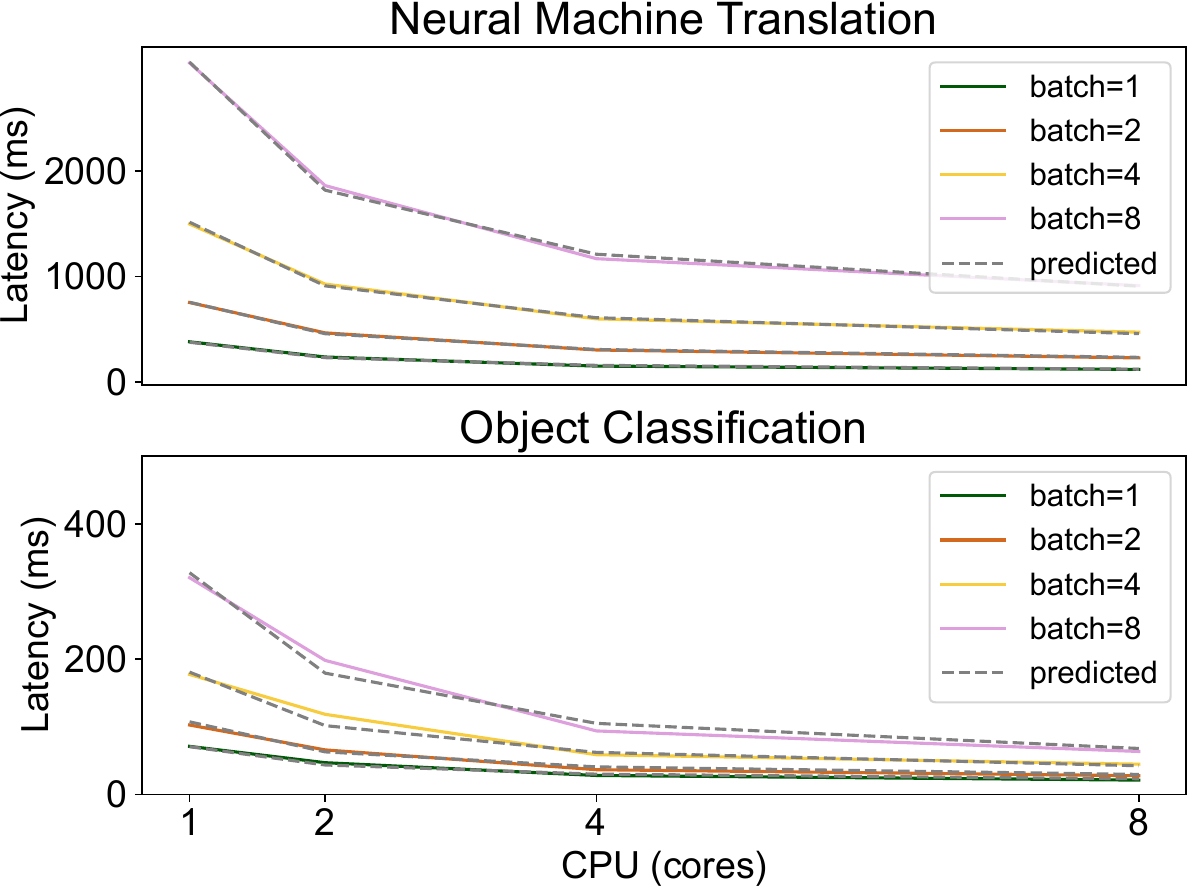}
    \caption{Performance profile of Translator and Classifier DL models with different CPU and batch configurations.}
    \label{fig:dl_profiling}
    % \vspace{-0.4cm}
\end{figure}

\noindent\textbf{DL Models}: To compare \namex{} with existing solutions in realistic environments, we use DL models across the computer vision, natural language processing, and audio recognition domains. These domains are also extensively used in other DL inference serving systems~\cite{2019-eurosys-grandslam, 2020-socc-inferline, romero2019infaas, romero2021llama, razavi2022fa2, ghafouri2024ipa}.
Table~\ref{tab:models_applications} summarises the DL models we use to evaluate \namex{}. We further assessed the equation in Formula~\ref{eq:latency_batch_core} with 512 requests per batch size and CPU core configuration as demonstrated in Figure~\ref{fig:dl_profiling} to show the effectiveness of the equation.

\begin{table}[!t]
\centering
\small
\caption{DL Models and Applications}
\subcaption*{DL Models}
\begin{tabular}{@{}p{3.3cm}rr@{}}
\toprule
\textbf{Task} & \textbf{Architecture} & \textbf{Abbreviation} \\
\midrule
Object Detection~\cite{yolov5} & YOLOv5n & OD\\
Object Classification~\cite{resnet18} & ResNet18 & OC \\
Audio to Text~\cite{wang2020fairseq} & FAIRSEQ S2T & AT\\
Sentiment Analysis~\cite{distillbert} & DistillBERT & SA \\
Language Identification~\cite{xlm-roberta} & XLM-RoBERTa & LI \\
Neural Machine Translation~\cite{elan-mt} & Elan-mt & NT \\
Text Summarization~\cite{t5-small} & T5-small & TS\\
\bottomrule
\end{tabular}
\bigskip
\subcaption*{Applications}
\begin{tabular}{@{}p{4.6cm}rr@{}}
\toprule
\textbf{Name} & \textbf{Pipeline} & \textbf{SLO} \\
\midrule
Video Monitoring & OD$\rightarrow$OC & 780 \\
Audio Sentiment Analysis & AT$\rightarrow$SA & 1350 \\
Natural Language Processing & LI$\rightarrow$NT$\rightarrow$TS & 2550 \\
\bottomrule
\end{tabular}
\label{tab:models_applications}
\vspace{-0.2cm}
\end{table}

\noindent\textbf{Pipelines}: We evaluate \namex{} using three pipelines as specified in Table~\ref{tab:models_applications} similar to the pipelines used in recent works~\cite{zhang2017live, clarifai, du2023rapsai, wu2022ai}. For the SLO of each pipeline, we follow a similar methodology described in~\cite{gujarati2017swayam} and calculate the processing latency with the least batch size and core allocation ($b=c=1$), aggregate the processing latencies of DL models, and multiply the result by a factor of three.

\noindent\textbf{Baselines}: We compare \namex{} to a state-of-the-art horizontal scaling (FA2~\cite{razavi2022fa2}) and an extended version of the recently proposed in-place vertical scaling (Sponge~\cite{razavi2024sponge}) in inference serving systems. FA2 solves the horizontal autoscaling problem by simultaneously considering the joint problem of batch size and number of instance selections. It increases the resource utilization of the DL models' instances, reducing the needed instances. Sponge uses in-place vertical scaling to react to sudden changes in the workload. However, their approach only considers one model in the system. To be able to compare \namex{} with Sponge, we use the approach in Algorithm~\ref{algo:dp_vertical} without the horizontal scaling part (one instance per DL model) as an extension of Sponge to solve the in-place vertical autoscaling problem in inference pipelines.

\noindent\textbf{Metrics}: We consider the following metrics in the evaluation, collected by Prometheus~\cite{prometheus}:
\begin{itemize}
    \item SLO violation rate: We collect each stage latency (consisting of queuing and processing latencies), aggregate them, and use them to check whether a request has violated the application's SLO.
    \item Cost: We use the number of instances $\times$ allocation CPU cores per instance of each DL model and report the aggregated results.
    \item Request dropping: We consider request dropping to avoid pending requests in the queues. We collect the dropped requests and use them to calculate the total SLO violation rate.
    \item P99 latency: Each second, multiple requests are being served. We pick the P99 of the requests to measure \namex{} and compare it with the P99 latency of the baselines.
\end{itemize}

\subsection{End-to-End Performance}
\begin{figure}
    \centering
    \includegraphics[width=0.95\linewidth]{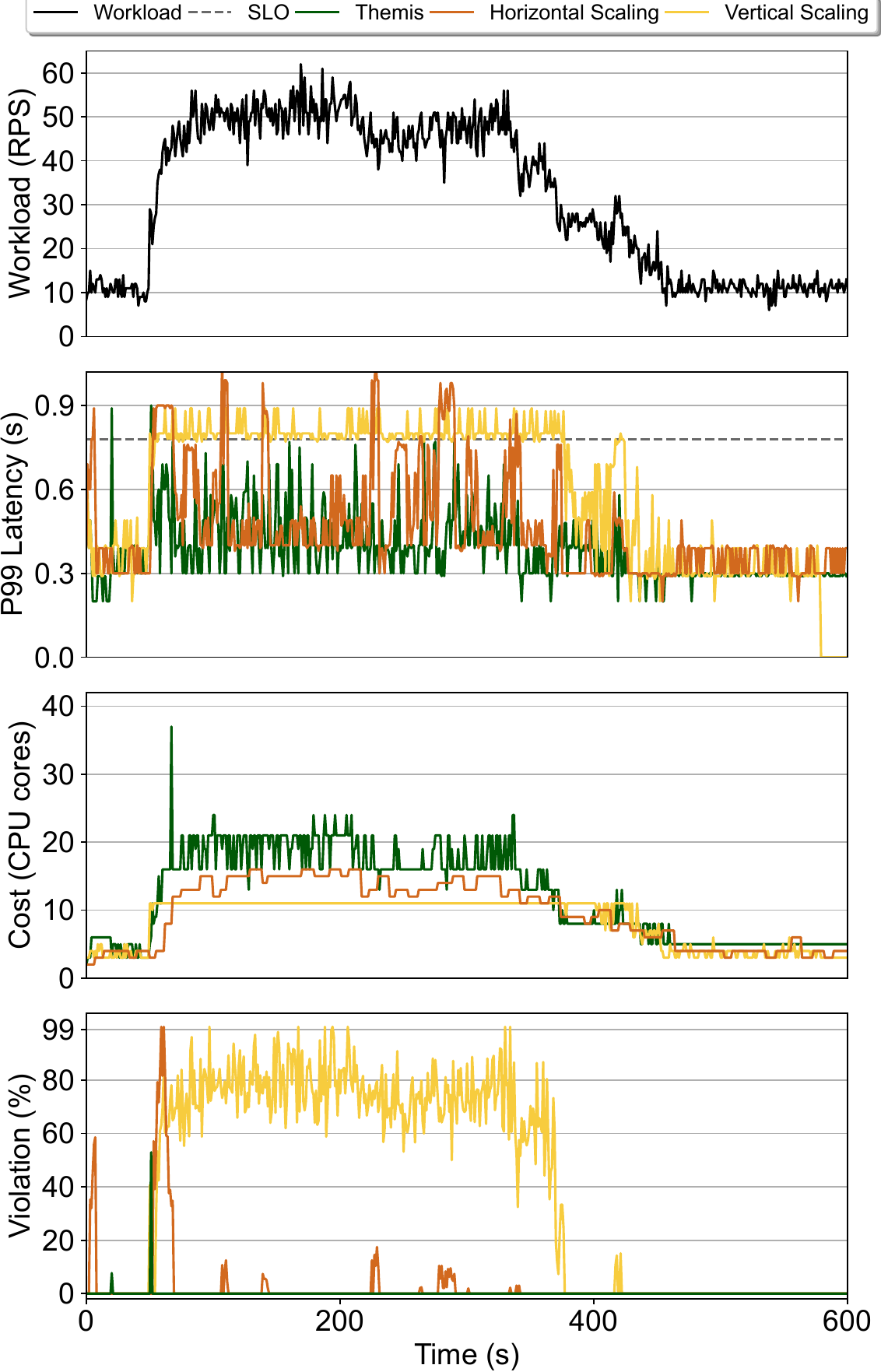}
    \caption{End-to-end comparison on the video monitoring pipeline. \namex{} reduces the SLO violation to roughly 0.1\% by using both scaling mechanisms jointly. Horizontal and vertical scaling mechanisms violate 2.4\% and 39.3\% of the requests' SLO, respectively.}
    \label{fig:e2e_evaluation_video}
    \vspace{-0.4cm}
\end{figure}

\noindent\textbf{Video Monitoring.} Figure~\ref{fig:e2e_evaluation_video} provides a detailed evaluation of \namex{}, horizontal scaling, and vertical scaling used to handle the Twitter trace workloads in the video monitoring pipeline. The top sub-figure shows the workload, starting low and stable, sharply increasing around the \nth{50} second, peaking near the \nth{180} second, and gradually returning to the initial RPS around the \nth{450} second.

When the workload starts, horizontal scaling violates more than half of the requests since the initial state with one instance on each DL model is not capable of supporting the incoming workload. To support the workload, horizontal scaling brings new instances up, where the cold start-up issue kicks in and starts violating the requests until the instances are spawned and warmed up. On the other hand, both \namex{} and vertical scaling immediately increase the number of CPUs for their running instances to increase their throughput, hence supporting the starting workload.

After a few seconds, in \namex{}, when the optimizer detects the workload is stable with the help of the LSTM, it switches to more 1-core instances to reduce the costs and, at the same time, to increase the DL models' throughput (second 10). The situation becomes complex when the workload increases to over 40RPS from 10RPS, where none of the approaches have enough up-and-running resources to capture the surge of requests, resulting in violating the SLOs. However, \namex{} and vertical scaling approaches change the computing resources of the running instances to the maximum to reduce the SLO violations. \namex{} has a slightly lower SLO violation rate since, in the previous seconds, it has brought up more 1-core instances to reduce the overall cost. The horizontal scaling approach initially has the highest violation rate since it needs to bring up new instances to support the current workload, but after the instances are up and running, it distributes the workload to multiple instances and stops violating the requests, which takes roughly 15 seconds. The vertical scaling approach adapts itself to the initial part of the workload. However, it starts suffering from SLO violation since one instance, even with the maximum resource allocation per DL model, is insufficient to support the incoming workload. After detecting the workload surge, \namex{} changes the computing resources of the running instances to maximum and simultaneously commands the system to bring up new instances to support the increased workload. With this approach, \namex{} violates the requests in less than a few seconds by compensating it with more cost at the \nth{55} second. After capturing the surge, the optimizer reduces the allocated resources to half to save costs by switching to horizontal scaling.
After that, there are some fluctuations in the workload where \namex{} is capable of capturing immediately due to the responsiveness of in-place vertical scaling, but horizontal scaling still shows lots of SLO violations (over 10\% in some seconds), where it needs to bring up new instances and violates SLOs meanwhile. After the workload reaches its initial RPS (\nth{450} second), all the approaches serve requests efficiently. 

\begin{figure}
    \centering
    \includegraphics[width=0.95\linewidth]{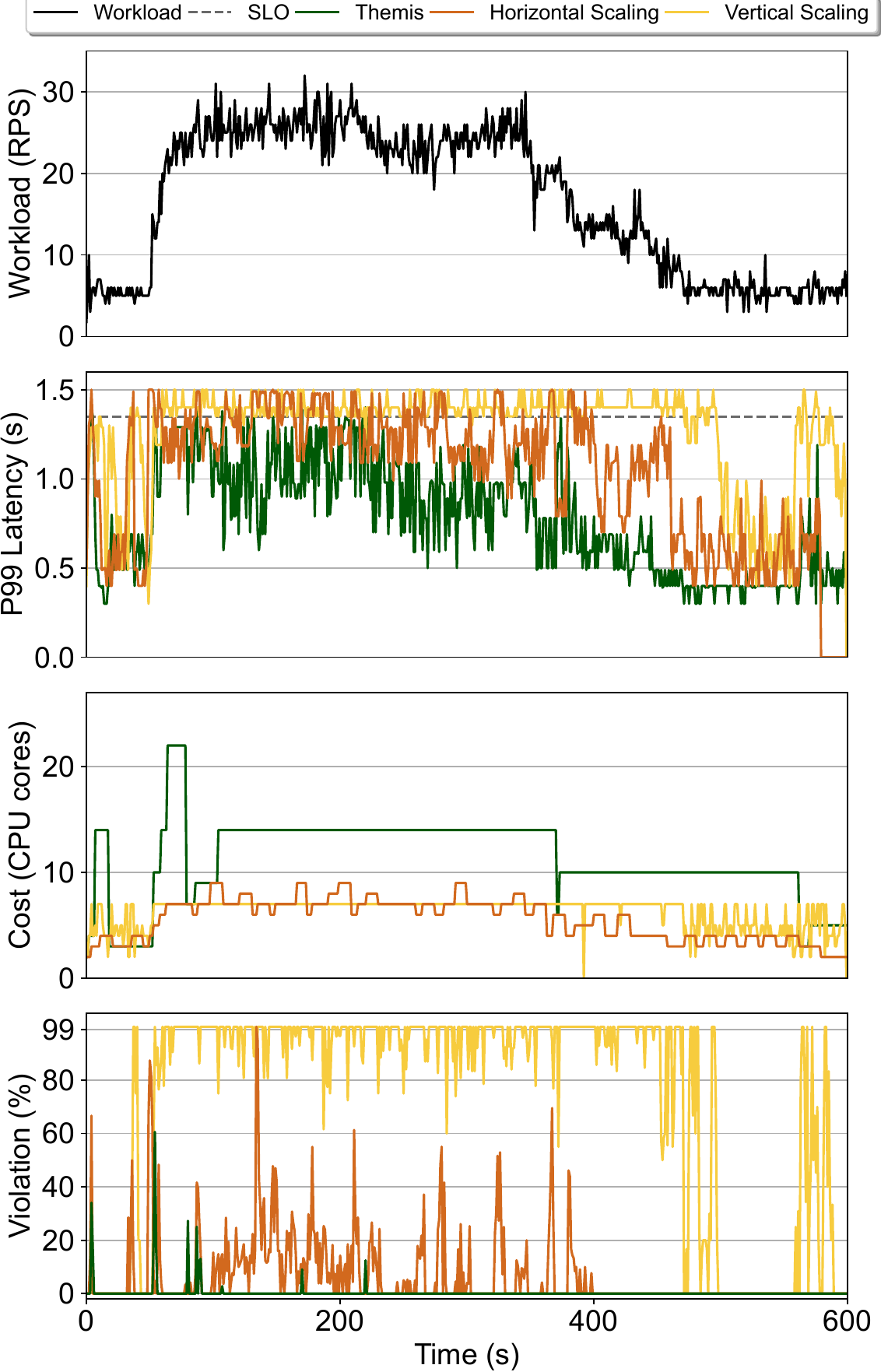}
    \caption{End-to-end comparison on the audio sentiment analysis pipeline. \namex{} reduces the SLO violation to less than 0.5\%. Horizontal scaling violates 6.5\% of the requests, while vertical scaling violates most of the requests, reaching over 72\% of the total SLO violation.}
    \label{fig:e2e_evaluation_sentiment}
    \vspace{-0.4cm}
\end{figure}

\noindent\textbf{Audio Sentiment Analysis.} Figure~\ref{fig:e2e_evaluation_sentiment} depicts the end-to-end evaluation over the audio sentiment analysis pipeline. 
Both models in the audio sentiment analysis pipeline are heavier than the ones in video monitoring, and with the same hardware, the system can process a lower RPS. Therefore, we use almost half of the previous workload in this experiment.

Unlike the video monitoring pipeline, the gap between the SLO and the processing latency with $b = c = 1$ is greater, and \namex{} uses the opportunity and increases the DL models' batch size more aggressively, resulting in higher throughput and lower needed resources. Because of the bigger batch sizes, the system does not need to switch to lower CPU cores using in-place vertical scaling, unlike the video monitoring pipeline, which changes the CPU cores more frequently to keep up with the workload. On the other hand, the bigger batch sizes increase the pipeline SLO violation from 0.1\% to 0.5\%.
The violation rate increases to 6.5\% in horizontal scaling due to the fact that the models are heavier, and the initialization takes longer, thereby a slower reaction time and a higher SLO violation rate. Vertical scaling suffers from the queuing back pressure throughout the experiment, demonstrating the importance of distributing the requests to multiple instances.
Also, at the moment, we depend on the LSTM and the performance profiles to switch to horizontal scaling to improve resource efficiency. Our LSTM might delay this process by predicting higher values for the maximum workload of the next period, as depicted in the same figure.

\begin{figure}
    \centering
    \includegraphics[width=0.95\linewidth]{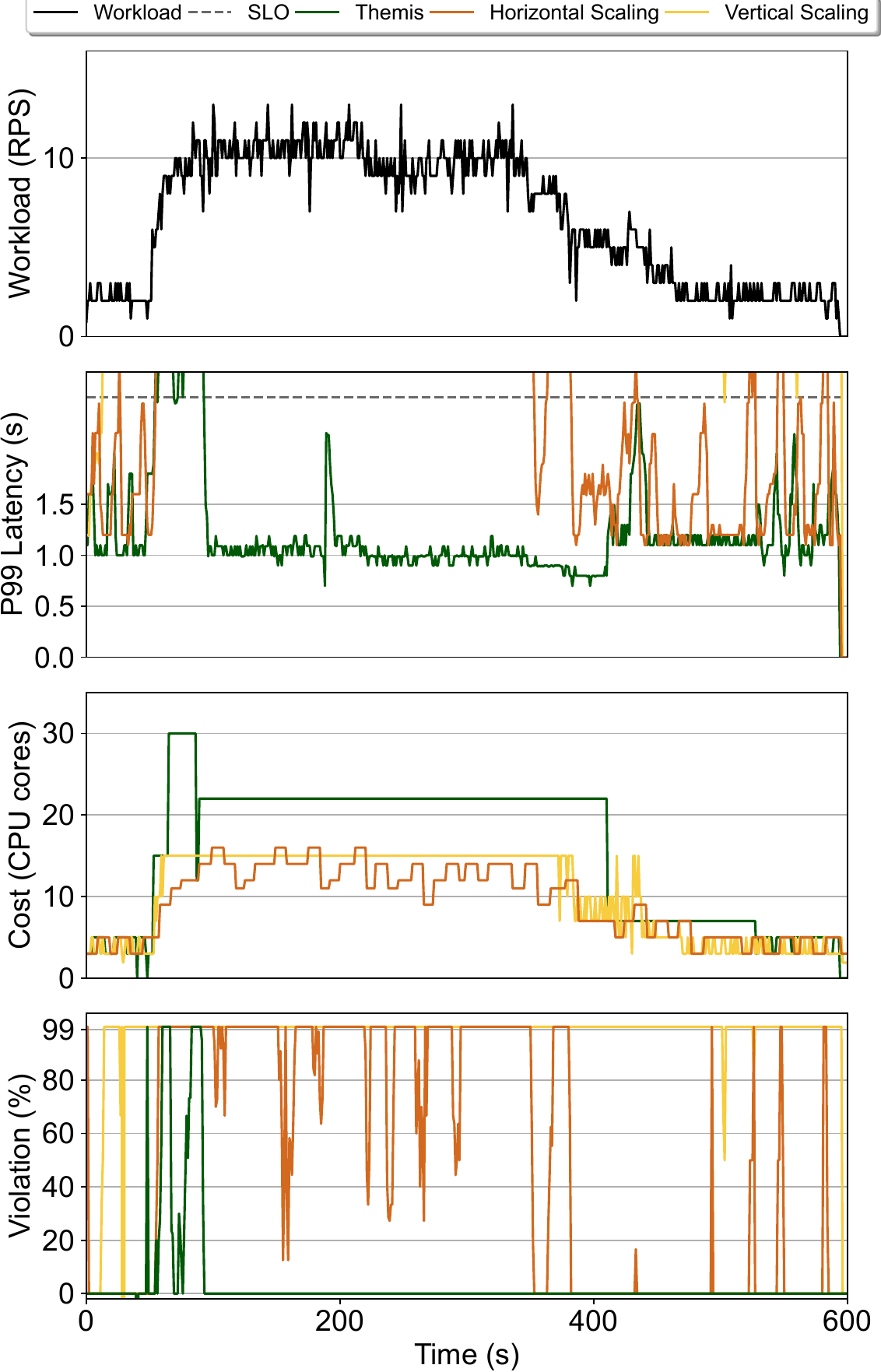}
    \caption{End-to-end comparison on the natural language processing pipeline. \namex{} reduces the SLO violation to less than 3.8\%. Horizontal scaling violates 50.7\% of the requests SLO, and vertical scaling violates 96.7\% of the requests SLO.}
    \label{fig:e2e_evaluation_nlp}
    \vspace{-0.3cm}
\end{figure}

\noindent\textbf{Natural Language Processing.} Figure~\ref{fig:e2e_evaluation_nlp} demonstrates the end-to-end evaluation over the natural language processing pipeline. The pipeline contains three DL models (compared to two DL models in the other pipelines), which are the most resource-intensive DL models in our experiments. Therefore, we scale the RPS to the range of $(2, 12)$. Following the same pattern, \namex{} scales the resources vertically at the beginning, and after capturing the initial workload, it reduces the resources to a minimum to match the workload. Horizontal scaling suffers the most in this scenario since the DL models are heavy and bring up new instances with the weight loading time taking over 10 seconds, resulting in violating roughly half of the requests SLO. Vertical scaling drops almost all the requests (over 96\%) due to not having enough resources.

Overall, \namex{} demonstrates adaptability to workload changes, effectively scaling resources up and down in response to demand. Despite workload spikes, the system manages to keep P99 latency mostly within the SLO threshold, indicating efficient latency management. The CPU usage graph reflects resource efficiency, increasing during high demand and conserving resources when demand is low. The figure highlights the effectiveness of these strategies in managing varying workloads, maintaining low latency and low cost, and minimizing SLO violations. 
% Finally, vertical scaling shows to be an asset in computer vision DL models inference serving (ResNet18 and YOLOv5n) due to their parallelizable architectures compared to the models in the other domains.

\subsection{Intra and Inter Parallelism}
In deep learning frameworks like TensorFlow~\cite{abadi2016tensorflow} and PyTorch~\cite{paszke2019pytorch}, two key parameters significantly impact request processing latency: \textit{inter} and \textit{intra} operation parallelism. These parameters have been extensively studied to determine their optimal configurations for various scenarios~\cite{zheng2022alpa}.

The \textit{intra} operation parallelism parameter allows tasks within a single batch of requests to be parallelized, thereby enhancing processing efficiency within that batch. This parameter can be dynamically adjusted during runtime, offering flexibility and adaptability to changing workloads. In contrast, the \textit{inter} operation parallelism parameter enables multiple tasks to be executed in parallel across different batches. However, unlike \textit{intra} operation parallelism, the \textit{inter} operation parameter is initialized once and remains fixed. This static nature can pose challenges, especially with in-place vertical scaling, where resource allocation needs may shift dynamically.

\begin{figure}
    \centering
    \includegraphics[width=0.95\linewidth]{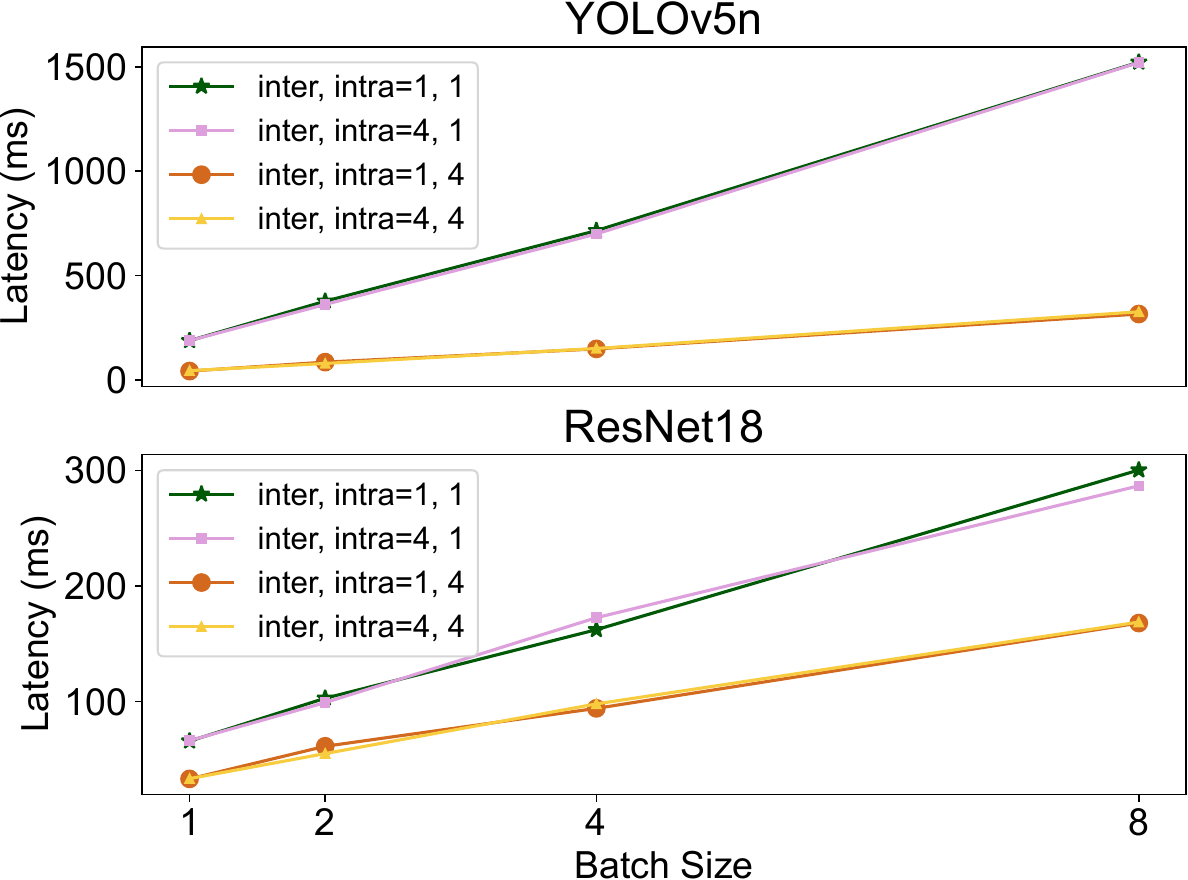}
    \caption{The effect of Intra and Inter parallelism parameters on the YOLOv5n and ResNet18 models with 4 CPU cores. Inter parallelism parameter does not affect the processing latency when there is one batch at a time in the DL model.}
    \label{fig:inter_intra}
    % \vspace{-0.2cm}
\end{figure}

As shown in Figure~\ref{fig:inter_intra}, the \textit{inter} operation parameter has minimal impact if new batches of requests are sent only after the previous batches have been fully processed. This suggests that the effectiveness of \textit{inter} operation parallelism is context-dependent, proving more beneficial in scenarios with concurrent batch processing rather than sequential processing.
%, which is the scenario in \namex{}.

\subsection{Request Dropping Effect}
When there is a surge of requests on the system, they may accumulate in the queues, leading to violations not only of their SLOs but also of new incoming requests, as older requests must be processed first. To prevent this, inference serving systems implement request dropping to alleviate queue back pressure~\cite{razavi2022fa2, ghafouri2024ipa}.

Different request-dropping strategies can be employed to minimize SLO violations. One strategy is to drop any request already reaching SLO at any processing or queuing stage. Another approach is to allow a buffer, such as three times the SLO, acknowledging that some delay might be acceptable. Finally, an approach is never to drop requests, opting to serve all requests regardless of the cost.

\begin{figure}
    \centering
    \includegraphics[width=0.95\linewidth]{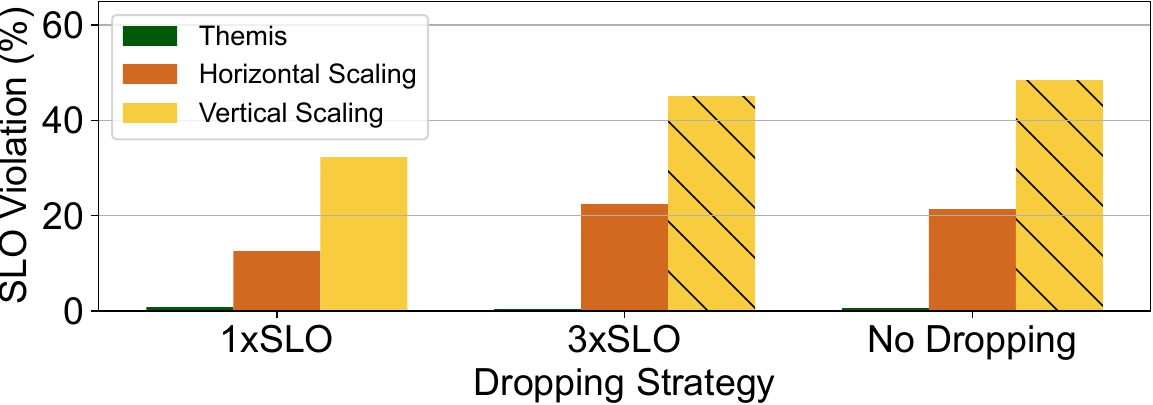}
    \caption{The effect of different dropping strategies. The 1xSLO dropping strategy helps reduce the total number of SLO violations.}
    \label{fig:dropping}
    % \vspace{-0.4cm}
\end{figure}

Figure~\ref{fig:dropping} compares these three strategies using \namex{} and baseline scaling approaches over the first 100 seconds of the workload depicted in Figure~\ref{fig:e2e_evaluation_video}, which includes a sudden increase in demand. \namex{} successfully serves almost all requests (over 99\%) during this period with virtually no SLO violations. In contrast, when the dropping strategy is relaxed (3x SLO and No Dropping), both horizontal and vertical scaling mechanisms experience significant request dropping, with over 20\% and 40\% drop in horizontal and vertical scaling, respectively. Thus, the 1x SLO strategy is the most effective in minimizing SLO violations for these approaches.

%% file: sections/sec_7_related_works.tex
\section{Related Work} \label{sec:relatedwork}
This section discusses autoscaling in inference serving systems and microservices since DL models are mainly encapsulated in microservices, and the inference pipelines are composed of several interconnected microservices as discussed in Section~\ref{sec:motivation}.

\noindent\textbf{Autoscaling in Inference Serving Systems}:
Multiple approaches have been proposed to reduce total resource consumption either using vertical or horizontal scaling in inference serving systems while guaranteeing SLOs~\cite{2020-socc-inferline, romero2019infaas, ghafouri2024ipa, romero2021llama, hu2021rim, 2020-osdi-serving, salmani2023reconciling, 2020-socc-gslice, 2019-sosp-nexus, vinod2022jellyfish, gunasekaran2022cocktail, 2017-nsdi-clipper}.
FA2~\cite{razavi2022fa2} uses graph transformation and dynamic programming techniques to reduce the number of instances in horizontal scaling. The graph transformation component deconstructs the execution graph, simplifying it and enabling optimal solutions to be found in real time. Meanwhile, the dynamic programming solution determines the optimal batch size and scaling factor for each DL model within the pipeline, ensuring efficiency and performance optimization.
Sponge~\cite{razavi2024sponge} uses a greedy approach to guarantee end-to-end latency of one DL model under a dynamic network using in-place vertical scaling. Similar to this work, they encapsulate the autoscaling problem into an IP and solve it in real time.

\noindent\textbf{Autoscaling in Microservices}: Current practices in microservice autoscaling mechanisms are mainly rule-based heuristics~\cite{kubernetes_hpa, kubernetes_vpa, 2016-icac-gru,2017-fgcs-cloudapp,2017-cascon-elascale,2019-www-smartvm, 2018-osdi-utune}, meta-heuristics~\cite{2017-tsc-autoscaling,2013-icse-genetic,2018-icdcs-rightsizing}, or recently, machine learning based~\cite{sachidananda2024erlang, rossi2019horizontal, qiu2023aware}.
Kubernetes Vertical Pod Autoscaler (VPA)~\cite{kubernetes_vpa} and Horizontal Pod Autoscaler (HPA)~\cite{kubernetes_hpa} manage the allocation of computing resources and the scaling of instances for microservices using threshold-based metrics. VPA adjusts the CPU and memory resources allocated to individual pods based on their current usage to ensure optimal performance. Conversely, HPA scales the number of instances up or down in response to real-time demands by monitoring metrics like CPU and memory utilization.
However, it is advised not to use VPA and HPA together, as they may interfere with each other's operations. When both are set, their concurrent adjustments can lead to conflicting actions, resulting in inefficient resource allocation and potential performance degradation.
A close work, SHOWAR~\cite{baarzi2021showar}, uses the empirical variance in the historical resource usage for vertical scaling to have an improved resource allocation compared to the default commonly used metric-based Kubernetes Vertical Autoscaler~\cite{kubernetes_vpa} (VPA), and uses CPU scheduler's eBPF metrics to design a more accurate horizontal autoscaler than Kubernetes Horizontal Autoscaler~\cite{kubernetes_hpa} (HPA). In vertical scaling in SHOWAR and Kubernetes VPA, the microservices are bounded by a maximum of 1-core CPU allocation since microservices do not leverage multiple CPU cores to accelerate the microservices' performance in contrast to DL models. Furthermore, the vertical autoscaler in SHOWAR does not react to changes in the workload but to CPU and memory metrics, which results in SLO violations in terms of unpredictable changes in the workload.

%% file: sections/sec_8_conclusion.tex
\section{Conclusion}
\label{sec:conclusion}
In this paper, we have presented \namex{}, a novel system designed to enhance inference serving through an innovative two-stage autoscaling strategy. By combining in-place vertical scaling with horizontal scaling, BoS effectively addresses the challenges of managing resource allocation under variable and unpredictable workloads. Extensive evaluation with real-world workloads demonstrates \namex{} reduces the SLO violation by over $10\times$ and minimizes the cost when the demand is low, showing \namex{}'s potential to provide responsive and cost-effective inference-serving solutions.

For future works, we consider enhancing \namex{}'s capability to utilize heterogeneous hardware resources effectively. Modern data centers often comprise a mix of CPUs, GPUs, TPUs, and specialized accelerators. Integrating \namex{} with hardware-aware scaling policies can optimize the allocation of tasks to the most suitable hardware, further improving efficiency and performance. 
% Further future direction includes deploying shared models across multiple applications and users. Future work must also investigate how \namex{} can support shared model deployments, ensuring efficient resource usage and reduced inference latency for all clients. This involves developing strategies for dynamic resource partitioning and load balancing that account for shared model access patterns and prioritization. 
Moreover, different placement strategies can be employed where new instances should reside. It requires answering the question of whether we should use a bin packing approach to utilize the physical machines to the maximum, a fair distribution of instances using a round-robin approach to not exhaust the machines, or if the most sparse instance distribution generates the ideal configuration in terms of guaranteeing SLOs.

%% file: sections/acknowledgement.tex
\section*{Acknowledgements} This work has been supported by Deutsche Forschungsgemeinschaft (DFG, German Research Foundation) – Project-ID 210487104 - SFB 1053.